\documentclass[pra]{revtex4}
\usepackage[dvips]{graphicx}
\usepackage{amsmath}
\usepackage{amssymb}

\begin{document}

\title{Charge induced energy fluctuations in thin organic films: effect on charge transport}

\author{Sergey V. Novikov}
\email{vanlab@online.ru}
\affiliation{A.N. Frumkin Institute of
Electrochemistry, Moscow, Russia}

\begin{abstract}
Effect of static charges on charge carrier transport in disordered
organic materials is considered. Long range nature of Coulomb
interaction requires to take into consideration a finite thickness
of the transport layer. Presence of conducting electrodes
significantly modifies properties of organic medium, removes a
long range Coulomb divergence, and makes it possible to calculate
in finite form statistical properties of organic medium (with
nonzero total charge density), relevant for transport
characteristics. A special attention is paid to the particular
case of charge induced disorder - a disorder originated from the
surface charge located at the rough surface of electrode. We
present also a generalization of 1D model of charge carrier
transport to the case of inhomogeneous energetic disorder that
realizes in for charge induced disorder.
\end{abstract}

\maketitle

\section{INTRODUCTION}

In recent years a significant attention has been paid to
experimental and theoretical study of the charge carrier transport
in disordered organic materials (for good reviews see Refs.
\cite{weiss,bassler,pope}). The obtained results profoundly
changed our understanding of the characteristics of the charge
carrier transport in these materials. The essence of a modern
paradigm of the charge carrier transport is a close connection
between spatial correlation of energy levels $U(\vec{r})$ of
transport sites in particular material and its transport
properties \cite{dpk,prl}. The mostly well understood reason for a
strong spatial correlation is a presence in the material some
particular kind of molecules, e.g. dipolar or quadrupolar
molecules \cite{dpk,prl,spie98}. The most important parameter of
the theory is the variance
$\sigma^2=\left<U^2\right>-\left<U\right>^2$ which gives a scale
of energy fluctuations in the material. For example, in dipolar
materials mobility has the following dependence on electric field
$E$ and temperature $T$
\begin{equation}
\mu =\mu _{0}\exp \left[ -A\left( \frac{\sigma}{kT} \right)%
^{2}+C_0\left(\left(\frac{\sigma}{kT}\right)^{3/2}-\Gamma \right)
\sqrt{\frac{eaE}{\sigma}}\right],
\label{CDM}
\end{equation}
where computer simulation for 3D model gives $A=9/25, C_0=0.78,
\Gamma=2$ \cite{prl}, while an exact solution for 1D case gives
$A=1, C_0=2, \Gamma=0$ \cite{dpk}. The particular kind of mobility
field dependence $\ln \mu \propto E^{1/2}$ (a Poole-Frenkel
dependence) is a result of slow decay of correlation function
$C(\vec{r}) =\left<U(\vec{r})U(0)\right> \propto 1/r$.
 This general theory describes well
all major features of photoinduced charge transport in disordered
organic materials.

Until now a very limited attention has been paid to the influence
of charges on the correlation properties of disordered organic
materials (the only exception is the paper \cite{spie97_a}). These
charges may be movable charges flowing through the sample or
static charges inserted into the sample, e.g. by the pulse of
accelerated electrons. Effect of charged particles on transport
properties was mainly considered using the framework of the
oversimplified Poole-Frenkel model of the isolated Coulomb trap.
This approach totally neglects collective effects which are
significant because of the long range nature of the Coulomb
potential and, thus, is limited to the case of extremely small
concentration of charged particles. This very long range nature of
Coulomb interaction requires to include into consideration the
finite thickness of transport layer. In the most typical
configuration transport layer consists of a film of organic
material bounded by highly conductive (metallic) electrodes. We
show that contribution of image charges created by these
electrodes significantly modifies properties of organic medium.
For example, this effect removes a notorious Coulomb divergence
and  makes possible to calculate in a finite form properties (mean
carrier energy, its variance, correlation function, etc) of
organic medium with nonzero total charge density. Hence, this
approach presents a first step in order to understand the
influence of finite concentration of carriers on the transport
properties of materials (beyond the effect of space charge).

In this paper we limit our consideration exclusively to the case
of a single charge carrier moving in the presence of static charge
background, thus avoiding a tremendous problem of solving a
self-consistent problem of the simultaneous motion of many
charges. Additionally, the problem of static charges presents a
significant interest by itself, because this very situation may
realize in some particular cases, for example in the case of
radiation-induced charge generation.

It will be shown also that if surface of the conducting electrodes
is rough, then surface charge accumulated under voltage applied
generates an additional disorder in the bulk of organic transport
layer. We will consider its influence on transport properties of
the device.

At present we have two major approaches to consider charge
transport in disordered organic materials. One is a computer
simulation of the hopping process, and another is an exact
solution for the equilibrium carrier mobility in a simplified 1D
model \cite{dpk}
\begin{equation}
\mu =\frac{\mu _{0}}{\epsilon \int\limits_{0}^{\infty }dy\exp
\left( -\epsilon y+\beta ^{2}\left[ C(0)-C(ay)\right] \right) },
\label{1D}
\end{equation}
where $\epsilon =ea\beta E$, $\beta =1/kT$, and $a$ is a minimal
charge-dipole separation. In the most thoroughly studied case -
charge transport in dipolar glasses - it was shown that these two
approaches give essentially the same results with not very
important differences for some numeric coefficients \cite{prl}.

For this reason we use in this paper the 1D model for evaluation
of the transport properties of the devices. Word "devices" instead
of "materials" is used on purpose, because for the extremely long
ranged Coulomb forces the relevant properties of the transport
layer (correlation function etc) depend on the size of the device
(on the thickness of transport layer) as well as on microscopic
properties of the transport medium. In this case a magnitude of
the disorder in some particular domain of transport layer is not a
constant but depends on the location of this domain. In other
words, charge induced disorder present a case of inhomogeneous
energy disorder. In this paper we develop a generalization of the
1D model for this new case.

\section{Charge transport in 1D inhomogeneous systems}

Until now very little is known about 3D charge carrier transport
in inhomogeneous random systems. At the same time, the 1D model
suggested in Ref.~\onlinecite{dpk} proved to be a useful tool in
revealing the dependence of a carrier mobility on electric field
and temperature \cite{prl}. For this reason it seems to be a
reasonable starting point to study charge carrier transport in the
framework of the 1D model.

We are going to find a stationary solution $c_s(z)$ for the
diffusion of a charge carrier in the presence of external electric
field $E$, random potential $U(z)$, a stationary source of
carriers located at $z=0$, and a perfect sink at $z=L$
\begin{equation} \frac{\partial
c}{\partial t}+\frac{\partial J}{\partial z} = I_0\delta(z),
\label{problem}
\end{equation}
\[
J(z,t)=-D \left[\frac{\partial c}{\partial
z}+\beta\left(\frac{\partial U}{\partial z}-eE\right) c\right],
\hskip10pt J(0,t)=0, \hskip10pt c(L,t)=0, \hskip10pt \beta=1/kT.
\]
The most important quantity to calculate is a stationary velocity
\begin{equation}
v=\frac{I_0 L}{\int_{0}^{L}dz c_s(z)}.
\label{vel}
\end{equation}
A direct solution of Eq. (\ref{problem}) gives
\begin{equation}
v=\frac{L D}{\int_{0}^{L}dz e^{-\beta[ U(z)-eEz]}\int_z^L dz'
e^{\beta[U(z')-eEz']}}.
\label{sol}
\end{equation}
Unfortunately, to perform the subsequent averaging of $v$ over
random field $U(z)$ is a difficult task. For this reason we will
perform the calculation of the average time $\left<t_L\right>$ for
the carrier to reach the opposite electrode, where $t_L=L/v$. One
may reasonably assume that functional dependences of
$\left<v\right>$ and $L/\left<t_L\right>$ on the relevant
parameters ($E, T,$ etc) should be the same. We have
\begin{equation}
\left<t_L\right>=\frac{1}{D}\int_{0}^{L}dx \int_z^L dz' e^{e\beta
 E(z-z')}\left<e^{\beta[U(z')-U(z)]}\right>.
\label{T_avg}
\end{equation}
If $U(z)$ is a Gaussian random field, then
\begin{equation}
\left<t_L\right>=\frac{1}{D}\int_{0}^{L}dz \int_z^L dz' e^{e\beta
E(z-z')} \exp
\left(\beta\left[\left<U(z')\right>-\left<U(z)\right>\right]
+\frac{1}{2}\beta^2\left<[U(z')-U(z)-\left<U(z')\right>+\left<U(z)\right>]^2\right>\right)
\label{T_avg_G}
\end{equation}
\[
=\frac{1}{D}\int_{0}^{L}dz \int_z^L dz' e^{e\beta E(z-z')} \exp
\left(\beta\left[\left<U(z')\right>-\left<U(z)\right>\right]+\frac{1}{2}\beta^2\left[C(z,z)+C(z',z')-2C(z,z')\right]\right),
\]
\[
C(z,z')=\left<U(z)U(z')\right>-\left<U(z)\right>\left<U(z')\right>.
\]

In the case of $L \rightarrow \infty$ and a homogeneous random
energy $U(z)$, when $\left<U(z)\right>={\rm const}$ and $C(z,z')$
depends on $z-z'$ only, this result exactly corresponds to the
result of Ref.~\onlinecite{dpk}. Hence, result (\ref{T_avg_G}) is
a direct generalization of the corresponding result of
Ref.~\onlinecite{dpk} to the case of inhomogeneous random
potential.

Transformation  to new coordinates $s=z+z', q=z-z'$ reveals an
important symmetry of Eq. (\ref{T_avg_G}) in the case when
$\left<U(z)\right>=0$
\begin{equation}
\left<t_L\right>=\frac{1}{2D}\left(\int_{0}^{L}ds \int_0^s dq
e^{e\beta Eq}g(s,q)+\int_{L}^{2L}ds \int_0^{2L-s} dq e^{e\beta
Eq}g(s,q)\right),
\label{t_sum}
\end{equation}
\[
\hskip10pt g(s,q)=
\exp\left\{\frac{1}{2}\beta^2\left[C\left(\frac{s+q}{2},\frac{s+q}{2}\right)+
C\left(\frac{s-q}{2},\frac{s-q}{2}\right)-
2C\left(\frac{s+q}{2},\frac{s-q}{2}\right)\right]\right\}.
\]
 Making an
additional transformation $s \Rightarrow 2L-s$ in the second
integral in Eq. (\ref{t_sum}) we have
\begin{equation}
\left<t_L\right>=\frac{1}{2D}\left(\int_{0}^{L}ds \int_0^s dq
e^{e\beta Eq}g(s,q)+\int_{0}^{L}ds \int_0^{s} dq e^{e\beta
Eq}g(2L-s,q)\right)
\label{t_sum_1}
\end{equation}
\[
=\frac{1}{2D}\int_{0}^{L}ds \int_0^s dq e^{e\beta
Eq}\left[g(s,q)+g(2L-s,-q)h(2L-s,q)\right].
\]
(note that $g(s,q)$ is an even function of $q$). The sum
$g(s,q)+g(2L-s,-q)$ in Eq. (\ref{t_sum_1}) (and, hence,
$\left<t_L\right>$) is invariant under the transformation $z
\Rightarrow L-z, z' \Rightarrow L-z'$, so
\begin{equation}
\left<t_L^{\rightarrow}\right>=\left<t_L^{\leftarrow}\right>,
\label{sym}
\end{equation}
where $\left<t_L^{\rightarrow}\right>$ corresponds to carriers
generated at $z=0$ and absorbed at $z=L$, and
$\left<t_L^{\leftarrow}\right>$ corresponds to the reverse process
(with the correspondent inversion of the electric field). We will
see that in spite of this symmetry the temporal dependence of the
photocurrent transient may be very different for these two cases.

\section{General properties of the charge-induced energy disorder}

\subsection{Potential energy, generated by random sources}

We consider the most simple model of the transport layer - a slab
of organic material (infinite in two dimensions) bounded at $z =
0,L$ by two parallel conducting electrodes having zero potential.
Organic medium is modeled by the regular lattice with lattice
constant $a$ and fraction $c$ of all sites occupied by point
charges. Since we are mostly interested in the case $c \ll 1$, so
the model of the perfect space lattice is not a significant
limitation. Lattice sites are also occupied by organic molecules
(of dipolar, quadrupolar nature etc), their random positions and
orientations produce an intrinsic energy disorder, which we
neglect in the most part of the paper, because it is independent
of the charge induced disorder, provides an additive contribution
to quantities in question (correlation functions etc), and its
properties have been extensively studied
\cite{dpk,prl,spie98,spie97_a,dieckmann,nv_94,young,nv_95,nv_96,nv_97,nv_97a}.

Let us consider a general case of a potential energy, generated by
random sources
\begin{equation}
U(\vec{r})=\sum_n \eta_n f(\vec{r},\vec{r}_n, \nu_n).
\label{sumU}
\end{equation}
Here source function $f(\vec{r},\vec{r}_n,\nu_n)$ describes the
contribution from a particular source to the total sum
(\ref{sumU}) and $\nu$ is an additional random parameter (or
parameters) that may affect the strength of the contribution
(e.g., it may be an orientation of dipole in the case of dipolar
sources). Note that summation in Eq. (\ref{sumU}) is performed
over all lattice points and $\eta_n=1$ if a real source is located
at the site $n$ and $\eta_n=0$ otherwise. In this paper we need to
calculate positional averages of two kinds
\begin{equation}
\left<\left<\eta_n A(\nu_n)\right>\right>=
\left<\eta_n\right>\left<A(\nu_n)\right>=c \left<A(\nu_n)\right>,
\label{avg_U}
\end{equation}
and
\begin{equation}
\left<\left<\eta_m\eta_n A(\nu_m)A(\nu_n)\right>\right>=
\left<\eta_n^2\right>\delta_{mn}\left<A^2(\nu_n)\right>+
\left<\eta_m\right>\left<\eta_n\right>(1-\delta_{mn})
\left<A(\nu_m)\right> \left<A(\nu_n)\right>
\label{avg_etaeta}
\end{equation}
\[
=c\delta_{mn}\left<A^2(\nu_n)\right>
+c^2(1-\delta_{mn})\left<A(\nu_m)\right>\left<A(\nu_n)\right>.
\]
Double brackets in the left part of Eqs. (\ref{avg_U}) and
(\ref{avg_etaeta}) denote a simultaneous averaging over positions
and parameter $\nu$.

\subsection{Source function and correlation function for point charges}

In the case of random potential energy generated by randomly
located point charges function $f(\vec{r},\vec{r}_n, \nu_n)$ is
proportional to the electrostatic potential of the charge located
at $\vec{r}=\vec{r}_n$ (here and later we assume that charge
carrier has a positive charge)
\begin{equation}
f(\vec{r},\vec{r}_n, \nu_n)=e\varphi(\vec{r},\vec{r}_n),
\label{f_charge}
\end{equation}
and $\varphi(\vec{r},\vec{r}_n)$ obeys a Poisson equation
\begin{equation}
\Delta \varphi=\frac{\partial^2 \varphi}{\partial x^2
}+\frac{\partial^2 \varphi}{\partial y^2}+\frac{\partial^2
\varphi}{\partial z^2}=-\frac{4\pi
e}{\varepsilon}\delta\left(\vec{r}-\vec{r}_n\right)
\label{Poisson}
\end{equation}
with zero boundary conditions
\begin{equation}
\left.\varphi\right|_{z=0}=\left.\varphi\right|_{z=L}=0.
\label{P_bound}
\end{equation}
Hence, $\varphi(\vec{r},\vec{r}_n)$ is proportional to the Green
function of the Laplace equation with the corresponding boundary
conditions
\begin{equation}
\varphi(\vec{r},\vec{r}_n)=-\frac{4\pi
e}{\varepsilon}G(\vec{r},\vec{r_n}).
\label{prop}
\end{equation}
Green function $G(\vec{r},\vec{r_n})$ is calculated in the
Appendix \ref{AppA} and has a form
\begin{equation}
G(\vec{r},\vec{r}_n)=-\frac{1}{\pi L}\sum_{s=1}^{\infty}\sin
\frac{s\pi z}{L}\sin \frac{s\pi z_n}{L} K_0\left(\frac{s\pi
\rho}{L}\right),
\label{Gfunc}
\end{equation}
where $K_0(x)$ is the Macdonald function and
$\rho^2=(x-x_n)^2+(y-y_n)^2$.

We consider here a slightly more general case when two types of
static charges are present: positive with fraction $c^+_n$ and
negative with fraction $c^-_n$ (we assume that these fractions may
vary with $n$). Note, that $c^+_n+c^-_n \le 1$. Average energy of
the charge carrier is
\begin{equation}
\left<U(\vec{r})\right>=-\frac{4\pi e^2}{\varepsilon}\sum_n
(c^+_n-c^-_n)G(\vec{r},\vec{r}_n),
\label{U}
\end{equation}
and the correlation function $C(\vec{r},\vec{r'})$ is
\begin{equation}
C(\vec{r},\vec{r'})=\left<U(\vec{r})U(\vec{r'})\right>-
\left<U(\vec{r})\right>\left<U(\vec{r'})\right>=\frac{16\pi^2
e^4}{\varepsilon^2}\sum_n
\left[c_n^++c^-_n-(c_n^+-c^-_n)^2\right]G(\vec{r},\vec{r}_n)
G(\vec{r'},\vec{r}_n).
\label{C}
\end{equation}

\subsection{Uniform distribution of charges}

As a simple example we provide a result for the case of uniform
charge distribution with $c_n^+,c^-_n=$ const. Here
\begin{equation}
\left<U(z)\right>=\frac{2\pi e^2 (c^+-c^-)}{\varepsilon
a^3}\hskip2pt z(L-z),
\label{Uuni}
\end{equation}
and
\begin{equation}
C(\vec{r},\vec{r'})=C(z,z',\rho)=\frac{8 e^4
\left[c^++c^--(c^+-c^-)^2\right] \rho}{\varepsilon^2
a^3}\sum_{s=1}^{\infty}\frac{1}{s}\sin\frac{\pi s
z}{L}\sin\frac{\pi s z'}{L}K_1\left(\frac{\pi s \rho}{L}\right),
\label{Cni}
\end{equation}
\[
\rho^2=(x-x')^2+(y-y')^2.
\]
In the particular case $\rho=0$
\begin{equation}
C(z,z',0)=\frac{4\pi e^4
\left[c^++c^--(c^+-c^-)^2\right]}{\varepsilon^2 a^3 L
}z_{-}(L-z_{+}), \hskip10pt z_{+}={\rm max}(z,z'), \hskip10pt
z_{-}={\rm min}(z,z').
\label{Cni_ro0}
\end{equation}

Equation (\ref{Cni_ro0}) shows that even in the case of uniform
charge distribution the resulting energetic disorder is not a
homogeneous one: the correlation function does depend on $z,z'$
separately, and not on the difference $z-z'$ only, as it should be
for the homogeneous case. Of course, the reason for this
phenomenon is an effect of conducting electrodes, which is
important for a very long range Coulomb interaction. Magnitude of
the disorder induced in a transport layer could be measured by the
variance $\sigma^2(z)=C(z,z)$. For the uniform distribution of
charges
\begin{equation}
\sigma^2(z)=\frac{4\pi e^4
\left[c^++c^--(c^+-c^-)^2\right]}{\varepsilon^2 a^3 L}z(L-z),
\label{sigma_uni}
\end{equation}
it attains a maximum at $z=L/2$ and goes to zero at the electrode
surfaces (as it should be, because electrodes have a constant
potential).

A direct calculation gives for the time (\ref{T_avg_G})
\begin{equation}
\left<t_L\right>=\frac{L^2}{D}\int_0^1 d\xi \int^1_\xi d\xi'
\exp\left[\lambda l S(\xi,\xi')\right],
\label{time_uni}
\end{equation}
\[
S(\xi,\xi')=-2\pi\lambda c_2 (\xi-\xi')^2 +2\pi l c_1
(\xi^2-\xi^{'2})+f(\xi-\xi'),
\]
\[
\lambda=\frac{\beta e^2}{\varepsilon a}, \hskip10pt l=\frac{L}{a},
\hskip10pt f=\frac{\varepsilon a^2 E}{e}-2\pi\lambda c_2-2\pi l
c_1, \hskip10pt c_1=c^+-c^-, \hskip10pt c_2=c^++c^--(c^+-c^-)^2.
\]
Introducing new coordinates $\zeta=\xi+\xi'$ and $\eta=\xi'-\xi$
we have
\begin{equation}
\left<t_L\right>=\frac{L^2}{2D}\int_0^1 d\eta \exp \left[ \lambda
l \left(-2\pi\lambda c_2 \eta^2 -f\eta \right)\right]
\int^{2-\eta}_{\eta} d\zeta \exp\left(-2\pi l^2\lambda c_1
\zeta\eta\right)
\label{time_uni_2}
\end{equation}
\[
=\frac{L^2}{2D \pi l^2\lambda c_1}\int_0^1 \frac{d\eta}{\eta} \exp
\left[ 2\pi\lambda^2 l c_2(\eta-\eta^2)-e\beta E L
\eta\right]\sinh\left[2\pi l^2\lambda c_1\left(\eta-
\eta^2\right)\right].
\]
One can see that Eq. (\ref{time_uni_2}) is an even function of
$c_1$, so charge carrier interacts in the same way with static
backgrounds created either by trapped electrons or holes. It is
easy to show that this particular property is valid for any charge
distribution having symmetry
$\left<U(z)\right>=\left<U(L-z)\right>$ (spatial symmetry of
$C(z,z')$ is not important because it is an even function of
$c_1$). We should emphasize that this particular property holds
for the Gaussian approximation (\ref{T_avg_G}) only.

Let us discuss Eq. (\ref{time_uni_2}) in the particular case
$c_1=0$, i.e. when the medium is neutral. This case was briefly
discussed in Ref.~\onlinecite{spie97_a} for the infinite medium.
We have
\begin{equation}
\left<t_L\right>_n=\frac{L^2}{D}\int_0^1 d\eta (1-\eta)\exp \left[
\lambda l \left(-2\pi\lambda c_2 \eta^2 -f\eta \right)\right]=
H(f)+\frac{1}{\lambda l}\frac{\partial H}{\partial f},
\label{time_uni_neutral}
\end{equation}
\[
H(f)=\frac{L^2}{D}\int_0^1 d\eta \exp \left[ \lambda l
\left(-2\pi\lambda c_2 \eta^2 -f\eta
\right)\right]=\frac{L^2}{2D\lambda\sqrt{2 l c_2}}
\exp\left(\frac{l f^2}{8 \pi c_2} \right)\left[{\rm erf}\left(
\frac{4 \pi\lambda l c_2+l f}{2\sqrt{2\pi l c_2 }}\right)-{\rm
erf}\left(\frac{l f}{2\sqrt{2\pi l c_2 }} \right) \right].
\]
Finally,
\begin{equation}
\left<t_L\right>_n=\frac{L^2}{4\sqrt{\pi} D\lambda^2 l c_2}
\left\{(A+B)e^{B^2}\left[{\rm erf}(A+B)-{\rm erf}(B)
\right]-\frac{1}{\sqrt{\pi}}\left(1-e^{-A^2-2AB}\right)\right\},
\label{time_uni_neutral_2}
\end{equation}
\[
A=\lambda\sqrt{2\pi l c_2}, \hskip10pt B=f\sqrt{\frac{l}{8\pi
c_2}}.
\]

The most prominent feature of Eq. (\ref{time_uni_neutral_2}) is
its behavior for $L\rightarrow \infty$. If $f > 0$  (i.e. if $E >
E_{cr}$), then $\left<t_L\right>_n \propto L$ and carriers have a
well defined average velocity which does not depend on $L$
\begin{equation}
\left<v\right>_{\infty}=e\beta D\left(E-E_{cr}\right), \hskip10pt
E_{cr}=\frac{2\pi\beta e^3 c_2}{\varepsilon^2 a^3}.
\label{Inf_L}
\end{equation}
If $f < 0$, then $\log\left<t_L\right>_n \propto L$ and average
carrier velocity goes to 0. Transition occurs at $E=E_{cr}$ in
full agreement with Ref.~\onlinecite{spie97_a}.

We would like to note a close connection between the case of
charge carrier transport in the infinite neutral medium with
$c^+=c^-$ and the anomalous diffusion of the particle in a 1D
random force field, described by the Langevin equation
\begin{equation}
\frac{dz}{dt}=\frac{1}{\gamma}F(z)+\eta(t),
\label{langevin}
\end{equation}
with a thermal noise $\eta(t)$
\begin{equation}
\left<\eta(t)\right>=0, \hskip10pt
\left<\eta(t)\eta(t')\right>=2\frac{kT}{\gamma}\delta(t-t'),
\label{noise}
\end{equation}
and a quenched random force $F(z)$ being a Gaussian white noise
\begin{equation}
\left<F(z)\right>=F_0, \hskip10pt
\left<F(z)F(z')\right>-F_0^2=\sigma\delta(z-z').
\label{F_noise}
\end{equation}
This problem was discussed thoroughly in
Refs.~\onlinecite{a_diff,a_diff_rev}, it demonstrates the same
transition from the mobile carriers with non-zero average velocity
to the immobile ones. The relation between the case $c^+=c^-$ (for
$L \rightarrow \infty$) and the problem (\ref{langevin}) arises
from the identical forms of the corresponding energy-energy
correlation functions. In both cases
$\left<\left[U(z)-U(z')\right]^2\right> \propto |z-z'|$. A more
careful analysis of this relationship will be studied elsewhere.

\begin{figure}
\begin{center}
\includegraphics[width=3.5in]{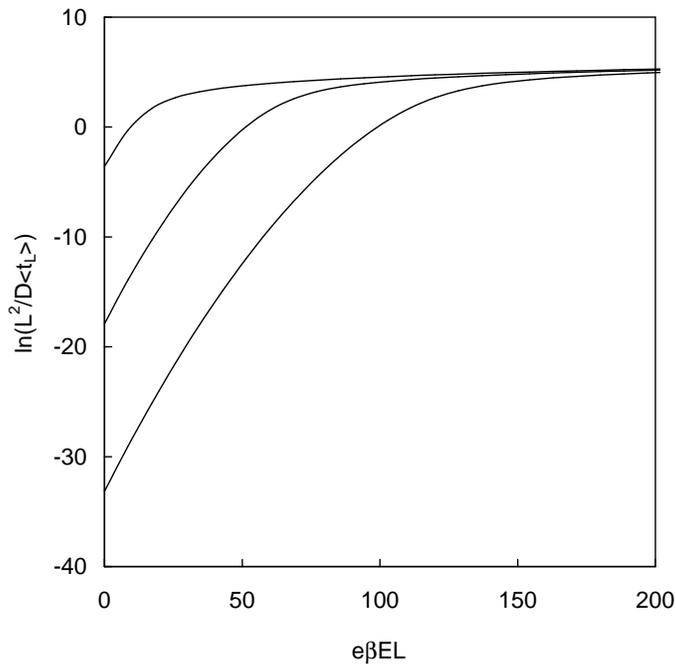}
\end{center}
\caption{Field dependence of $\left<t_L\right>$ for the case of
uniformly charged background with $L/a=150$, $e^2
\beta/\varepsilon a = 20$, $c^-=0$, and different values of
$c^{+}$: $1 \times 10^{-5}$, $3 \times 10^{-5}$, and $5 \times
10^{-5}$, from upper curve downwards, correspondingly.}
\label{fig1}
\end{figure}

In the general case of macroscopically charged medium with $c_1
\ne 0$ we can estimate time (\ref{T_avg_G}) for strong disorder,
when $2\pi l^2\lambda |c_1| \gg 1$. In this case we can retain
only the leading exponential term in (\ref{time_uni_2}), so
\begin{equation}
\left<t_L\right>\approx\frac{L^2}{2D l^2\lambda |c_1|}
\sqrt{\frac{2\lambda l (\lambda c_2+l|c_1|)}{2\pi\lambda l
(\lambda c_2+l|c_1|)-e\beta
EL}}\hskip2pt\exp\left[\frac{\left(2\pi\lambda l (\lambda
c_2+l|c_1|)-e\beta EL\right)^2}{8\pi \lambda l (\lambda
c_2+l|c_1|)}\right].
\label{spoint_1}
\end{equation}
This expression is valid if
\[
2\pi\lambda l (\lambda c_2+l|c_1|)-e\beta EL \gg \left[2\pi\lambda
l (\lambda c_2+l|c_1|)\right]^{1/2}.
\]
Note, that if $c_1 \ne 0$, then $\log\left<t_L\right> \propto L^2$
for $L \rightarrow \infty$, so for a infinite charged medium there
is no non-dispersive transport. Field dependence of time
(\ref{time_uni_2}) is shown in Fig. \ref{fig1}.

The case of the uniform distribution of static charges does
realize in the case of radiation-induced charge generation in
disorder organic materials. In this particular case movable
charges are induced in the material by the pulse of accelerated
electrons. In the most typical case electrons have high energy, so
concentration of electron-hole pairs (usually, electrons get
trapped and holes are movable) created in the bulk of the material
is approximately constant and does not decay while going away from
electrode. We could expect that the model of uniformly charged
material is a reasonable model for description of transport
properties in the radiation-induced generation, at least for large
times, when concentration of movable charges is much smaller than
the concentration of static charges.

A different case of static charges trapped at the close vicinity
of the electrode (this case does realize for the radiation-induced
generation of charge carriers by the pulse of low energy
electrons) will be studied elsewhere.

\subsection{How valid is the Gaussian approximation?}

Particular case of the uniform charge distribution gives us a good
possibility to discuss a validity of the Gaussian approximation
for the 1D transport model. Gaussian approximation is valid for
the main body of the distribution if $\sigma^2(z) \gg U_{s}^2$,
where $U_{s}=e^2c^{1/3}/\varepsilon a$ is a typical contribution
from the single charge source, thus ensuring that the resulting
potential energy is not dominating by the single contribution.
This inequality leads to
\begin{equation}
\frac{4\pi c^{1/3}}{a L}z(L-z) \gg 1.
\label{valid_Gauss}
\end{equation}
Hence, the Gaussian approximation is valid in the bulk of the
layer if $\pi c^{1/3} L/a \gg 1$, but in the near vicinity of
electrode at  $z \le z_b$, $z_b \simeq a/4\pi c^{1/3}$
non-Gaussian effects are significant. If $4\pi c^{1/3} \ge 1$ ($c
\ge 5 \times 10^{-4}$), then for the whole transport layer the
main body of distribution has a Gaussian shape.

Nevertheless, a Gaussian shape for {\it the tail} of the
distribution holds only if a more strong inequality is valid:
$\sigma^2(z) \gg U^2_{max}$, where $U_{max}=e^2/\varepsilon a$ is
a maximal contribution from the single source. This leads to
\begin{equation}
\frac{4\pi c}{a L}z(L-z) \gg 1,
\label{valid_Gauss_tail}
\end{equation}
so tail of the distribution is Gaussian in the bulk of the layer
if $\pi c L/a \gg 1$, and at the vicinity of electrodes if $4\pi c
\ge 1$ (or $c \ge 0.1$). The last condition seems to be difficult
to match.

\section{Random potential energy, generated by charged rough
metallic surface}

\subsection{Basic formalism}

\begin{figure}
  \begin{minipage}[t]{0.47\linewidth}
     \begin{center}
        \includegraphics[width=2in]{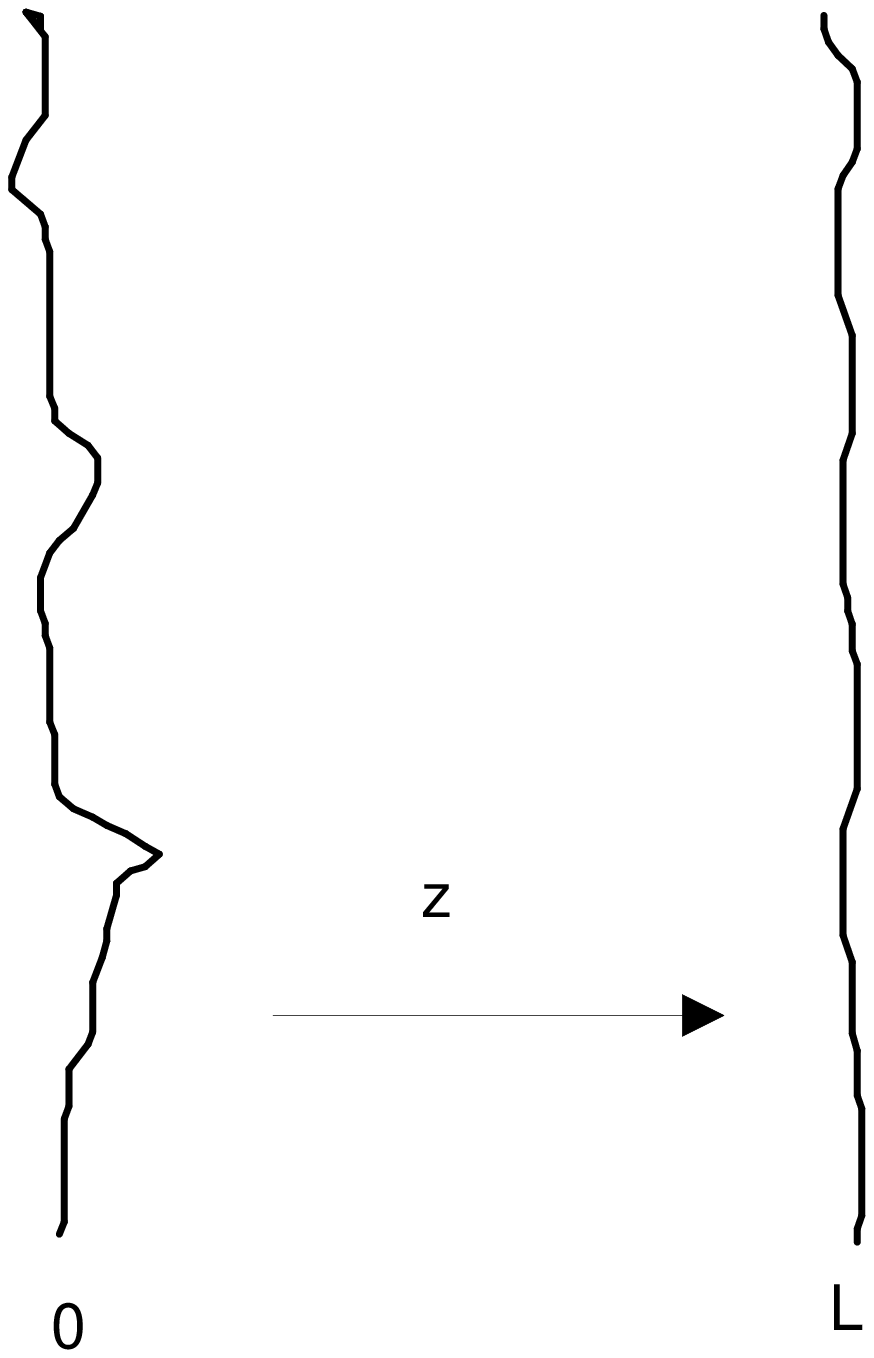}
\newline   (a)
     \end{center}
   \vspace{10pt}
  \end{minipage}%
  \hspace{20pt}
  \begin{minipage}[t]{0.47\linewidth}
     \begin{center}
        \includegraphics[width=3.2in]{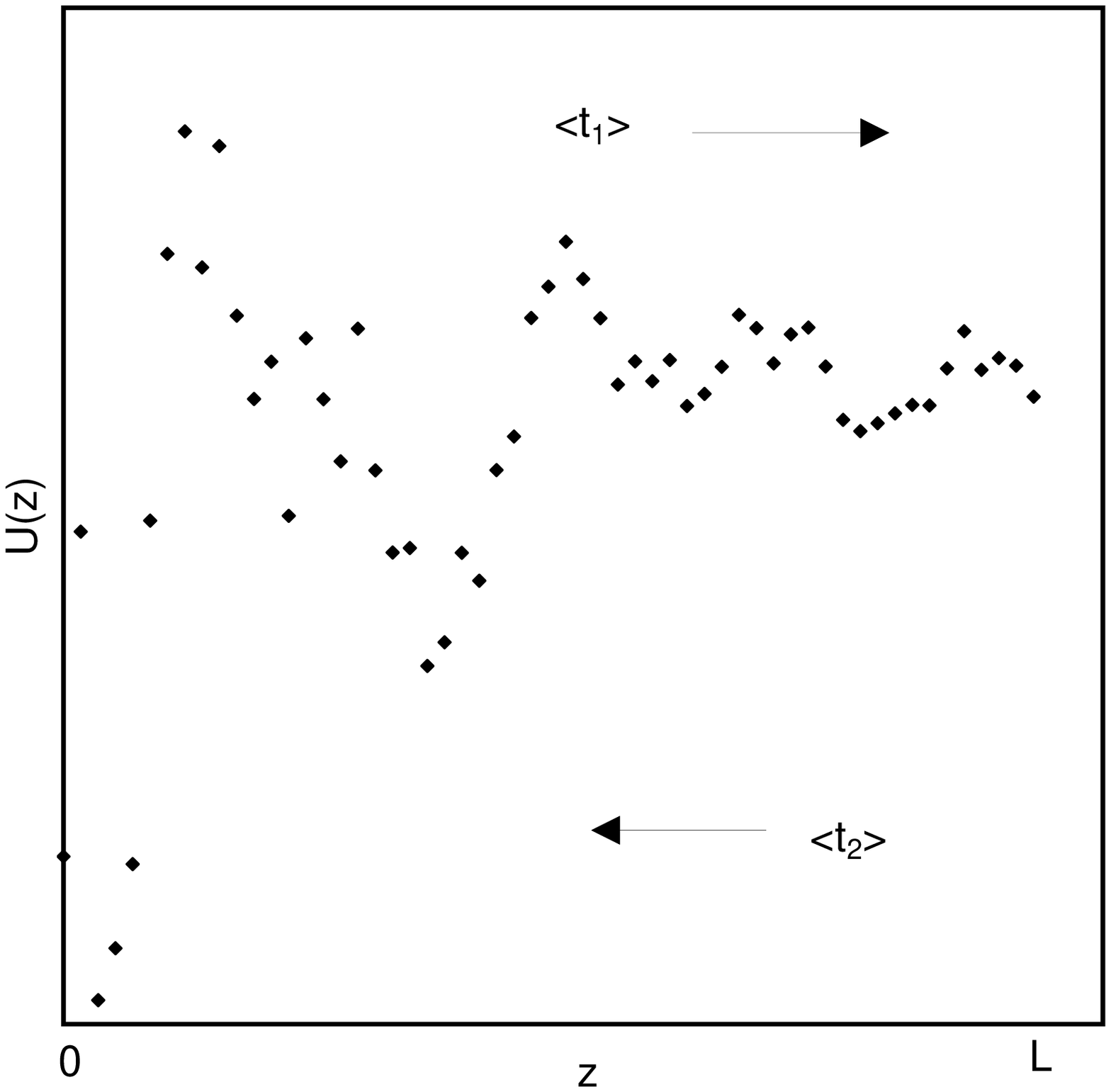}
\newline   (b)
     \end{center}
  \end{minipage}%
\newline
  \begin{minipage}[t]{0.47\linewidth}
     \begin{center}
        \includegraphics[width=3.2in]{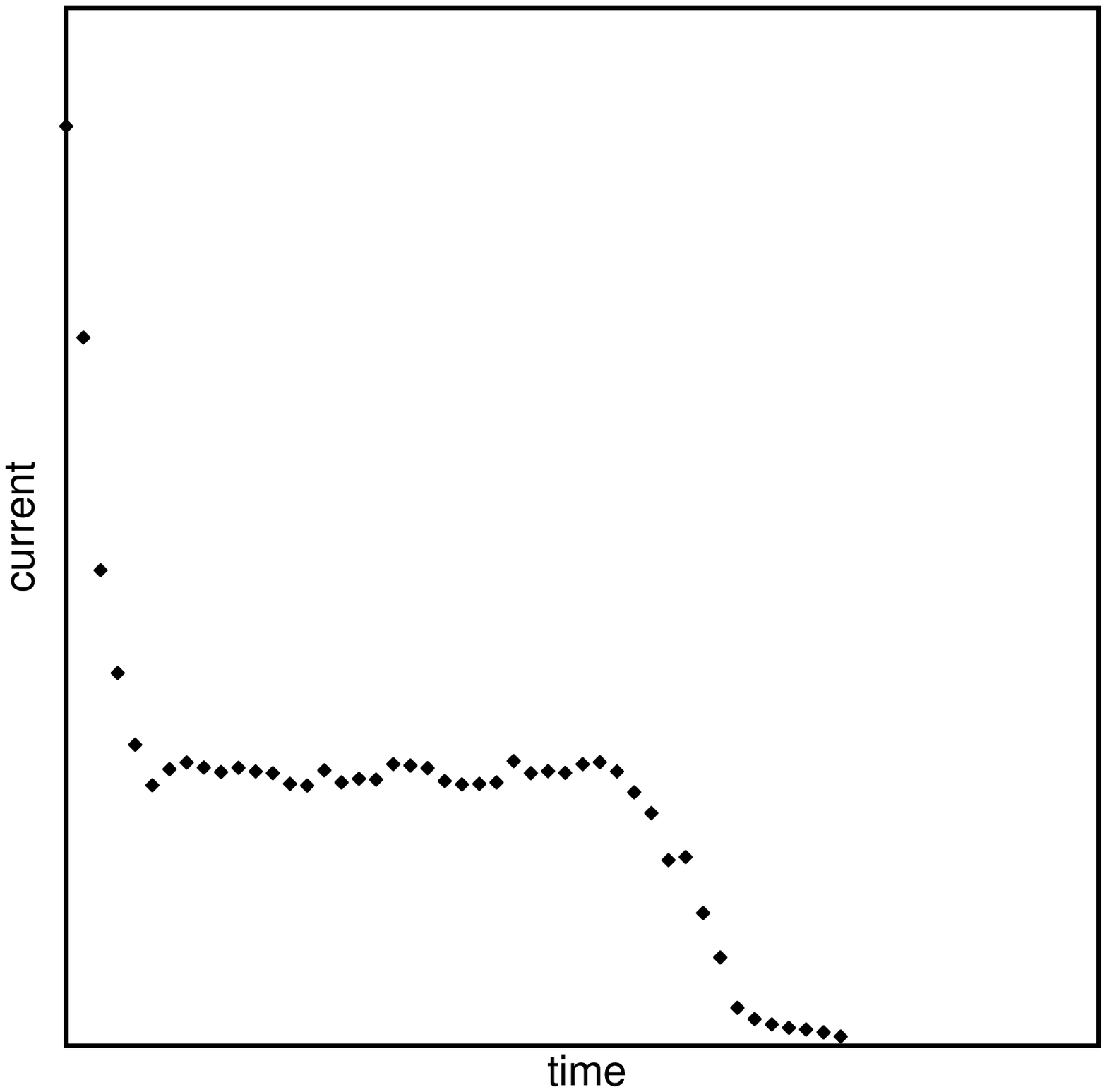}
\newline   (c)
     \end{center}
   \end{minipage}%
  \hspace{10pt}
  \begin{minipage}[t]{0.47\linewidth}
     \begin{center}
        \includegraphics[width=3.2in]{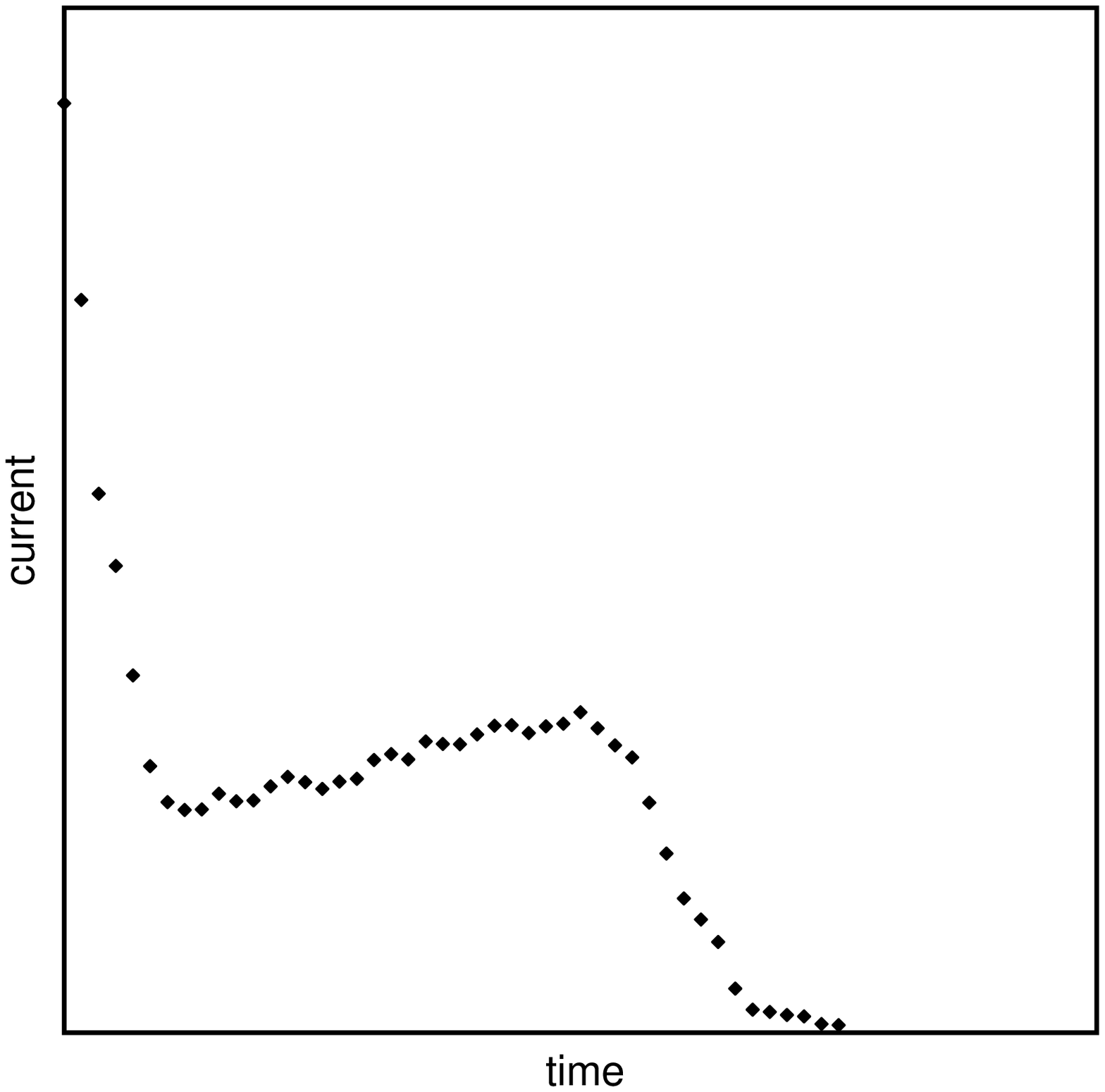}
\newline   (d)
     \end{center}
   \vspace{10pt}
  \end{minipage}%
  \hspace{20pt}
\caption{(a) Transport layer bounded by rough electrodes (note the
different roughness of the left and right electrode). (b) A
typical distribution of the random energy $U(z)$ in the bulk of a
transport layer induced by the rough surface of the electrodes in
the case of significant difference in their roughness. According
to Eq. (\ref{sym}) $\left<t_1\right>=\left<t_2\right>$, but time
dependence of the photocurrent may be quite different for these
two cases. (c,d) Typical time dependence of the photocurrent in
the time-of-flight experiment when a smooth electrode serves as an
injector (c), or a rough one serves as an injector (d).}
   \label{fig2}
\end{figure}

A typical transport device can be described as a flat capacitor
with the region between conducting electrodes filled by an organic
material (transport layer with the thickness $L$). If voltage
$V_0$ is applied to the capacitor, then surfaces of the electrodes
develop a surface charge with the density $Q$ proportional to the
electric field $E=-V_0/L$, namely $Q=E/4\pi$. In the ideal case of
absolutely flat electrodes' surfaces the surface charge just
maintains this very field $E=$ const in the bulk of organic
material, according to the well known formula for a correspondent
distribution of the electrostatic potential $\varphi$ in the bulk
of the capacitor
\begin{equation}
\varphi(\vec{r})=\int_S \frac{QdS}{R},
\label{Q}
\end{equation}
where integration is carried out over the electrode surfaces and
$R$ is a distance between point $\vec{r}$ and a particular part of
the surface. Situation is different in the case of a rough surface
of the electrodes (Fig. \ref{fig2}a). In this case we can describe
surface of the electrode as a random surface, so the distance $R$
in Eq. (\ref{Q}) fluctuates, thus the resulting potential
$\varphi$ is a random function. Hence, a roughness of the
electrode surface induces an additional energetic disorder in the
bulk of the organic transport layer. We are going to estimate a
magnitude of this kind of disorder and its effect on the transport
properties of devices. One could immediately note two important
properties of this kind of disorder: its magnitude increases with
the increase of the electric field $E$ applied to the device, and
decreases while going away from the electrodes (so this is another
case of inhomogeneous disorder).

Let us calculate the potential distribution inside a capacitor
having rough metallic electrodes. Surface of one electrode obeys
an equation $z=h_0(x,y)$, and another one $z=L+h_L(x,y)$, where
$h_0(x,y)$ and $h_L(x,y)$ are random functions and $h_0,h_L \ll
L$. We define the locations of the average electrode planes in
such a way that $\left<h_{0,L}\right>=0$. We are going to study
variation of the potential for $z \gg h$.

Inside the capacitor we should solve a Laplace equation
\begin{equation}
\Delta \varphi=\frac{\partial^2 \varphi}{\partial x^2
}+\frac{\partial^2 \varphi}{\partial y^2}+\frac{\partial^2
\varphi}{\partial z^2}=0
\label{Laplace1}
\end{equation}
with potential $\varphi$ obeying the boundary conditions
\begin{equation}
\left.\varphi\right|_{z=h_0(x,y)}=0, \hskip10pt
\left.\varphi\right|_{z=L+h_L(x,y)}=V_0.
\label{bound1}
\end{equation}
We are going to find a correction for the potential distribution
inside a capacitor which is resulted from the roughness of the
electrode surfaces
\begin{equation}
\varphi(\vec{r})=-Ez+\delta\varphi(\vec{r}).
\label{correction}
\end{equation}
Calculation of the first order correction to the potential
distribution with $\delta\varphi \propto h$ is performed in
Appendix \ref{AppB}.

\begin{figure}
\begin{center}
\includegraphics[width=3.5in]{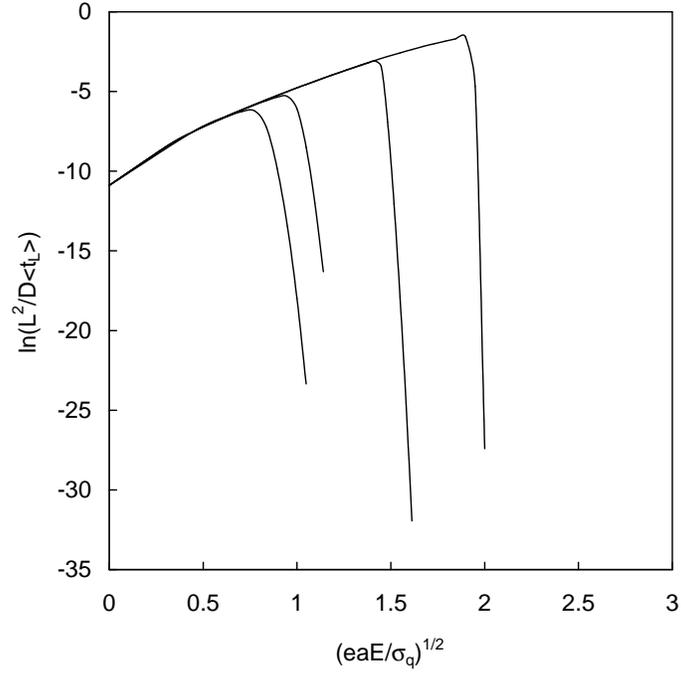}
\end{center}
\caption{Field dependence of $\left<t_L\right>$ for $L/a=100$,
$\left<h_0^2\right>^{1/2}/a=5$, $\left<h_L^2\right>^{1/2}=0$, in
the case of quadrupolar glass with $\sigma_q\beta=2$ bounded by
rough metallic electrodes with different values of $l_0/a$: 50,
25, 10, 5, from the right curve to the left, correspondingly.
Surface correlation function was chosen in the form
$\Omega_0(\vec{\rho})=\exp\left(-\rho^2/2 l_0^2\right)$, which is
suitable for a naturally rough metal surface \cite{rasigni1}. If
$a=1$ nm and $\sigma_q=0.1$ eV, then $eaE/\sigma_q \approx 1$ for
$E=1 \times 10^6$ V/cm.}
\label{fig3}
\end{figure}

The energy-energy correlation function has a form (note that
$\left<\delta \varphi(\vec{r})\right>=0$)
\begin{equation}
C(\vec{r_1},\vec{r_2})=e^2\left<\delta \varphi(\vec{r_1}) \delta
\varphi(\vec{r_2})\right>
\label{Cp}
\end{equation}
\[
= \frac{1}{4\pi^2}\int
d\vec{k}e^{i\vec{k}(\vec{\rho_1}-\vec{\rho_2})}\frac{1}{\sinh^2
kL}\left[U^2_0 \Omega_0(\vec{k})\sinh k(L-z_1)\sinh k(L-z_2)+U^2_L
\Omega_L(\vec{k})\sinh kz_1 \sinh kz_2\right],
\]
\[
U_j=eE \left<h_j^2\right>^{1/2}, \hskip10pt
\Omega_j(\vec{\rho_1}-\vec{\rho_2})=\frac{1}{\left<h^2_j\right>}
\left<h_j(\vec{\rho_1})h_j(\vec{\rho_2})\right>, \hskip10pt
\Omega_j(\vec{k})= \int
d\vec{\rho}e^{-i\vec{k}\vec{\rho}}\Omega_j(\vec{\rho}), \hskip10pt
j=0,L.
\]
(here $\vec{k}$ and $\vec{\rho}=(x,y)$ are 2D vectrors). We can
simplify equation (\ref{Cp}) in a situation when the surface
correlation function $\Omega(\vec{\rho})$ has a single length
scale $l$ (surface correlation length). In a very common situation
$l \gg \left<h^2\right>^{1/2}$ (this is so, for example, for the
so called naturally rough metal surfaces
\cite{rasigni1,rasigni2}). If $z \ll L$, then significant domain
in integral (\ref{Cp}) is $kl \sim 1$. If $z_1,z_2,l \ll L$ (we
consider a vicinity of one electrode), then we can simplify
equation (\ref{Cp})
\begin{equation}
C(\vec{r_1},\vec{r_2}) \approx \frac{U^2_0}{4\pi^2}\int
d\vec{k}\Omega_0(\vec{k})\exp(i\vec{k}(\vec{\rho_1}-\vec{\rho_2})-k(z_1+z_2)),
\label{Cp_dz_1}
\end{equation}
so if $z_1,z_2 \gg l$, then
\begin{equation}
C(\vec{r_1},\vec{r_2}) \propto U^2_0 l^2\int
d\vec{k}\exp(i\vec{k}(\vec{\rho_1}-\vec{\rho_2})-k(z_1+z_2))=
\frac{U^2_0 l^2 (z_1+z_2)}{\left[(z_1+z_2)^2+\rho^2\right]^{3/2}}.
\label{Cp_dz_2}
\end{equation}
This universal behavior takes place when a large area $S \gg l^2$
of the electrode gives a contribution to the distribution of the
potential. Such regime is observed for $z_1,z_2 \gg l$ only,
because essential area $S \propto z^2$. Note, that in this very
case we should expect that the distribution of $\delta\varphi$ has
approximately a Gaussian form (because many independent areas of
the surface give contributions to the resulting distribution). In
the opposite case, when $z_1,z_2 \ll l$, then
$C(\vec{r_1},\vec{r_2})\approx U_0^2$.

\subsection{What should we expect to observe in experiments?}

We can calculate time $\left<t_L\right>$ for the case of rough
electrodes using Eq. (\ref{T_avg_G}). For the most interesting
case $l \gg \left<h^2\right>^{1/2}$ results are shown in Fig.
\ref{fig3}. The most important feature of the mobility dependences
is an abrupt decrease of the carrier velocity when $E$ reaches a
critical value. Using a saddle-point method one can show that
$E_{\rm crit} \propto l/e\beta \left<h^2\right>$.

To the best of our knowledge, such abrupt drop of the mobility has
not been observed in experiments. Quite possibly, $E_{\rm crit}$
is so high that it cannot be reached because of the dielectric
breakdown of the layer. If $l=15$ nm, $\left<h^2\right>^{1/2}=1.8$
nm (data from Ref.~\onlinecite{rasigni1}), then $E_{\rm crit}
\approx 2 \times 10^6$ V/cm. This field is indeed high, but it
could be reached for layers of high quality (see
Ref.~\onlinecite{spg}). Another possible reason of the discrepancy
could be a failure of the 1D model in this particular case.
Finally, our approach totally neglects a correlation between rate
of the charge carrier injection from some electrode's area and its
 profile. We may expect that our approach is valid
for rather "smooth" rough surfaces, where we cannot expect a
significant variation of the injection rate over electrode
surface. Anyway, we believe that a careful study of charge
transport in high quality layers with simultaneous control (by the
use of scanning tunneling microscope) of the roughness of the
electrodes could confirm (or disprove) our results.

We should expect that a significant effect of the roughness could
be observed for weaker electric field if we consider a temporal
behavior of the photocurrent $I(t)$ for the time-of-flight
experiments. At present we cannot produce any analytic or
simulation results for $I(t)$. Anyway, we may safely suppose that:

\noindent 1) Transport should be more dispersive in the case of
rough electrodes in comparison to the case of smooth electrodes,
just because of the increase of disorder. Additionally, this
increase should lead to the decrease of mobility.

\noindent 2) If a rough electrode serves as an injector, we should
expect an unusual transformation of the temporal behavior of the
photocurrent $I(t)$ with the increase of the electric field $E$.
Typical distribution of sites' energies in this case is shown in
Fig. \ref{fig2}b. In a weak field region we should expect a
transient of the usual kind (Fig. \ref{fig2}c). In strong field
region, with the increase of the magnitude of the disorder near
the injecting electrode, transient should transform to the form
shown in Fig. \ref{fig2}d. A reason for the development of the
region where current increases with $t$ is carriers' acceleration
when they move to the region with smaller disorder. This kind of
photocurrent transformation with the increase of $E$ was indeed
observed in Ref.~\onlinecite{jap}.

\noindent 3) We noted previously a principal symmetry of the
equation (\ref{T_avg_G}), so
$\left<t_L^{\rightarrow}\right>=\left<t_L^{\leftarrow}\right>$
even when electrodes have very different degree of roughness
(compare the left and right electrode in Fig. \ref{fig2}a). But if
we consider a temporal behavior of the current $I(t)$, it could be
very different for these two cases. For not so small $E$, if a
smooth electrode serves as injector, we should expect a usual form
of $I(t)$ (see Fig. \ref{fig2}c), but if a rough electrode serves
as injector, we should expect the form shown in Fig. \ref{fig2}d.

\section{Conclusions}
We suggested a new approach for calculation of statistical
properties of disordered organic materials with nonzero charge
density. Explicit consideration of the finite transport layer
bounded by conducting electrodes removes a long range Coulomb
divergence and gives finite values for the relevant
characteristics of the device (correlation function etc). We
presented a generalization of the 1D model of the charge carrier
transport in disordered organic matrices to the case of
inhomogeneous disorder. This approach permitted us to calculate
transport properties of matrices with charge induced disorder. A
similar approach permits us to calculate an energy-energy
correlation function for the transport layer bounded by rough
conducting electrodes. We expect that the roughness of the
electrode surface should lead to the unusual temporal behavior of
the photocurrent transient (appearance of the region where current
increases with time) in the case of strong electric field.

\acknowledgments

This research was partially sponsored by the International Science
and Technology Center grant 872 and the Russian Fund for Basic
Research grant 99-03-32111.

\appendix
\section{Calculation of the Green function for the laplace equation}
\label{AppA}

Taking into account a geometry of the problem, we may write
\begin{equation}
G(\vec{r},\vec{r_n})=\frac{1}{4\pi^2}\int d\vec{k}
e^{i\vec{k}\vec{\rho}}G_k(z,z_n),
\label{Green}
\end{equation}
where $\vec{k}$ and $\vec{\rho}=(x-x_n,y-y_n)$ are 2D vectors, and
function $G_k(z,z_n)$ obeys an equation
\begin{equation}
\frac{d^2G_k}{dz^2}-k^2G_k=\delta(z-z_n), \hskip10pt
G_k(0,z_n)=G_k(L,z_n)=0, \hskip10pt G_k(z,z_n)=G_k(z_n,z).
\label{Green_Z}
\end{equation}
A direct calculation gives
\begin{equation}
G_k(z,z_n)=-\frac{\sinh kz_{-}\sinh k(L-z_{+})}{k \sinh kL},
\hskip10pt z_{+}={\rm max}(z,z_n), \hskip10pt z_{-}={\rm
min}(z,z_n).
\label{Gk_sol}
\end{equation}

We would like to note another useful expression for $G_k(z,z_n)$
which is based on the series of the eigenfunctions  of the
operator $\partial^2-k^2$
\begin{equation}
G_k(z,z_n)=\sum_{s=1}^{\infty}\frac{\psi_s(z)\psi_s(z_n)}{\lambda_s},
\hskip10pt \lambda_s=-k^2-\left(\frac{\pi s}{L}\right)^2 ,
\hskip10pt \psi_s(z)=\sqrt{\frac{2}{L}}\sin\left(\frac{\pi
sz}{L}\right).
\label{Gs_sol_ef}
\end{equation}
From Eq. (\ref{Gs_sol_ef}) it immediately follows that
\begin{equation}
\varphi(\vec{r},\vec{r}_n)=\frac{4e}{\varepsilon
L}\sum_{s=1}^{\infty}\sin \frac{s\pi z}{L}\sin \frac{s\pi z_n}{L}
K_0\left(\frac{s\pi \rho}{L}\right),
\label{fi_sol}
\end{equation}
where $K_0(x)$ is the Macdonald function. Last expression is
useful to show behavior of $\varphi(\vec{r},\vec{r}_n)$ for $\rho
\gg L$. Presence of metallic electrodes transforms the usual
Coulomb decay to the exponential one
\begin{equation}
\varphi(\vec{r},\vec{r}_n)\approx \frac{4e}{\varepsilon L} \sin
\frac{\pi z}{L}\sin \frac{\pi z_n}{L}
\sqrt{\frac{L}{2\rho}}\exp\left(-\frac{\pi
\rho}{L}\right),\hskip10pt \rho \gg L,
\label{fi_sol_asm}
\end{equation}
thus removing a nasty long range Coulomb divergence.

\section{First order correction to the potential distribution for
a transport layer bounded by rough electrodes}
\label{AppB}

To solve Eq. (\ref{Laplace1}) let us make a transformation of
variables
\begin{equation}
X=x, \hskip10pt Y=y, \hskip10pt Z=\frac{L(z-h_0)}{L+h_L-h_0},
\label{newZ}
\end{equation}
so the boundary conditions (\ref{bound1}) transform to
\begin{equation}
\left.\varphi\right|_{Z=0}=0, \hskip10pt
\left.\varphi\right|_{Z=L}=V_0.
\label{bound2}
\end{equation}
In new variables $(X,Y,Z)$ the Laplace equation (\ref{Laplace1})
has a form
\begin{equation}
\frac{\partial^2 \varphi}{\partial X^2 }+\frac{\partial^2
\varphi}{\partial Y^2}+\frac{\partial^2 \varphi}{\partial
Z^2}\left[\left(\frac{\partial Z}{\partial x
}\right)^2+\left(\frac{\partial Z}{\partial y
}\right)^2+\left(\frac{\partial Z}{\partial z
}\right)^2\right]+2\left(\frac{\partial^2 \varphi}{\partial X
\partial Z } \frac{\partial Z}{\partial x }+\frac{\partial^2 \varphi}{\partial Y
\partial Z } \frac{\partial Z}{\partial y }\right)+\frac{\partial \varphi}{\partial Z }
\left(\frac{\partial^2 Z}{\partial x^2 }+\frac{\partial^2
Z}{\partial y^2 }\right)=0.
\label{Laplace2}
\end{equation}
We will seek a solution of Eq. (\ref{Laplace2}) by the
perturbation theory (a formal small parameter being $h$)
\begin{equation}
\varphi=\sum_n \varphi_n, \hskip10pt \varphi_n \thicksim O(h^n),
\hskip10pt \varphi_0=V_0 Z/L, \hskip10pt
\left.\varphi_n\right|_{Z=0}= \left.\varphi_n\right|_{Z=L}=0,
\hskip10pt n \ge 1 .
\label{perturb}
\end{equation}
Our main goal is to calculate a leading term for the correlation
function $C(\vec{r_1},\vec{r_2})=e^2\left<\delta\varphi(\vec{r_1})
\delta\varphi(\vec{r_2})\right>$. For this reason we should
calculate $\varphi_1$ only.

Equation for an every term $\varphi_n$ has a general structure
\begin{equation}
\Delta \varphi_n=J_n,
\label{eq_n}
\end{equation}
where the source $J_n$ depends on $\varphi_k$, $0 \le k \le n-1$.
Solution of Eq. (\ref{eq_n}) is
\begin{equation}
\varphi_n(\vec{r})=\int
d\vec{r_1}G(\vec{r},\vec{r_1})J_n(\vec{r_1}),
\label{eq_n_sol}
\end{equation}
here $G(\vec{r},\vec{r_1})$ is the Green function (\ref{Green})
for the Laplace operator with zero boundary conditions
(\ref{perturb}).

The source term for the first order correction is
\begin{equation}
J_1=\frac{V_0}{L}\left[\left(1-\frac{Z}{L}\right)\Delta_\bot
h_0+\frac{Z}{L}\Delta_\bot h_L\right], \hskip10pt
\Delta_\bot=\frac{\partial^2}{\partial X^2
}+\frac{\partial^2}{\partial Y^2},
\label{corr1}
\end{equation}
so
\begin{equation}
\varphi_1(\vec{r})=\frac{V_0}{L}\int
d\vec{r_1}G(\vec{r},\vec{r_1})
\left[\left(1-\frac{Z_1}{L}\right)\Delta_\bot
h_0(\vec{\rho_1})+\frac{Z_1}{L}\Delta_\bot
h_L(\vec{\rho_1})\right].
\label{corr1a}
\end{equation}
Calculation of the integral (\ref{corr1a}) gives
\begin{equation}
\varphi_1(\vec{r})=-\frac{V_0}{4\pi^2 L}\int d\vec{k} \int
d\vec{\rho_1} \int_0^L dZ_1
e^{i\vec{k}(\vec{\rho}-\vec{\rho_1})}k^2
G_k(Z,Z_1)\left[\left(1-\frac{Z_1}{L}\right)h_0(\vec{\rho_1})+
\frac{Z_1}{L}h_L(\vec{\rho_1})\right]
\label{corr1_sol}
\end{equation}
\[
=-\frac{V_0}{4\pi^2 L}\int d\vec{k} e^{i\vec{k}\vec{\rho}}\int_0^L
dZ_1 k^2 G_k(Z,Z_1)\left[\left(1-\frac{Z_1}{L}\right)h_0(\vec{k})+
\frac{Z_1}{L}h_L(\vec{k})\right]
\]
\[
=\frac{V_0}{4\pi^2 L}\int d\vec{k}e^{i\vec{k}\vec{\rho}}
\left[\left(1-\frac{Z}{L}-\frac{\sinh k(L-Z)}{\sinh kL}
\right)h_0(\vec{k})+\left(\frac{Z}{L}-\frac{\sinh kZ}{\sinh
kL}\right) h_L(\vec{k})\right].
\]
The total first order correction to the potential distribution
(here we are going back to the original coordinates) is
\begin{equation}
\delta \varphi =\varphi-\left<\varphi\right>
=-\frac{V_0}{L}\left[\left(1-\frac{z}{L}\right)h_0(x,y)
+\frac{z}{L}h_L(x,y)\right]
\label{delta_phi}
\end{equation}
\[
+\frac{V_0}{4\pi^2 L}\int d\vec{k}e^{i\vec{k}\vec{\rho}}
\left[\left(1-\frac{z}{L}-\frac{\sinh k(L-z)}{\sinh kL}
\right)h_0(\vec{k})+\left(\frac{z}{L}-\frac{\sinh kz}{\sinh
kL}\right) h_L(\vec{k})\right]
\]
\[
=-\frac{V_0}{4\pi^2 L}\int d\vec{k}e^{i\vec{k}\vec{\rho}}
\left[\frac{\sinh k(L-z)}{\sinh kL} h_0(\vec{k})+\frac{\sinh
kz}{\sinh kL} h_L(\vec{k})\right].
\]

\end{document}